\documentclass[amsmath,amssymb,aps,prd,11pt,tightenlines,superscriptaddress,
nofootinbib,preprintnumbers,notitlepage]{revtex4-1}

\usepackage{multirow}
\usepackage{graphicx}
\usepackage{tabularx}
\usepackage{mathtools}
\usepackage{slashed}
\usepackage{booktabs}
\usepackage{url}
\usepackage[maxfloats=100]{morefloats}
\usepackage{color,xcolor}
\usepackage[normalem]{ulem}
\usepackage{subfig}
\usepackage[colorlinks=false]{hyperref}
\definecolor{dark-red}{rgb}{0.8,0.0,0.0}
\definecolor{dark-blue}{rgb}{0.0,0.0,0.8}
\definecolor{dark-green}{rgb}{0.,0.6,0.}
\usepackage{todonotes} 

\makeatletter
\def\l@subsubsection#1#2{}
\makeatother

\usepackage{enumitem}
\setlist[itemize]{itemsep=1pt,parsep=1pt, topsep=1pt}
\newcommand{\MGMCatNLO}{MadGraph5\_aMC@NLO} 
\newcommand{\pythia}{{\sc Pythia}}
\newcommand{\delphes}{{\sc Delphes}}
\newcommand{\fastjet}{{\sc FastJet}}
\newcommand{\met}{\ensuremath{\slashed{E}}}
\newcommand{\fbinv}{\mbox{\ensuremath{~\mathrm{fb^{-1}}}}}
\newcommand{\abinv}{\mbox{\ensuremath{~\mathrm{ab^{-1}}}}}

\newcommand{\kt}{\ensuremath{k_\mathrm{T}}}
\newcommand{\pt}{\ensuremath{p_\mathrm{T}}}

\newcommand{\ptfl}{\ensuremath{p_{\mathrm{T},4\ell}}}

\begin{document}


\title{Longitudinally polarized $ZZ$ scattering at the Muon Collider} 

\author{Tianyi Yang}
\affiliation{Department of Physics and State Key Laboratory of Nuclear Physics and Technology, Peking University, Beijing, 100871, China}

\author{Sitian Qian}
\affiliation{Department of Physics and State Key Laboratory of Nuclear Physics and Technology, Peking University, Beijing, 100871, China}

\author{Zhe Guan}
\affiliation{Department of Physics and State Key Laboratory of Nuclear Physics and Technology, Peking University, Beijing, 100871, China}

\author{Congqiao Li}
\affiliation{Department of Physics and State Key Laboratory of Nuclear Physics and Technology, Peking University, Beijing, 100871, China}

\author{Fanqiang Meng}
\affiliation{Department of Physics and State Key Laboratory of Nuclear Physics and Technology, Peking University, Beijing, 100871, China}

\author{Jie Xiao}
\affiliation{Department of Physics and State Key Laboratory of Nuclear Physics and Technology, Peking University, Beijing, 100871, China}

\author{Meng Lu}
\affiliation{
School of Physics, Sun Yat-Sen University, Guangzhou 510275, China}

\author{Qiang Li}
\affiliation{Department of Physics and State Key Laboratory of Nuclear Physics and Technology, Peking University, Beijing, 100871, China}

\date{\today}

\begin{abstract}
Measuring longitudinally polarized vector boson scattering in, e.g., the $ZZ$ channel is a promising way to investigate the unitarization scheme from the Higgs and possible new physics beyond the Standard Model. However, at the LHC, it demands the end of the HL-LHC lifetime luminosity, 3000\fbinv, and advanced data analysis technique to reach the discovery threshold due to its small production rates. Instead, there could be great potential at future colliders. In this paper, we perform a Monte Carlo study and examine the projected sensitivity of longitudinally polarized $ZZ$ scattering at a TeV scale muon collider. We conduct studies at 14~TeV and 6~TeV muon colliders respectively and find that a 5 standard deviation discovery can be achieved at a 14~TeV muon collider, with 3000\fbinv\ of data collected. While a 6~TeV muon collider can already surpass HL-LHC, reaching 2 standard deviations with around 4\abinv\ of data. The effect from lepton isolation and detector granularity is also discussed, which may be more obvious at higher energy muon colliders, as the leptons from longitudinally polarized Z decays tend to be closer.
\end{abstract}

\maketitle
\tableofcontents

\section{Introduction}
\label{sec:intro}

The discovery~\cite{plb:2012gu,plb:2012gk} and property measurements~\cite{higprop} of the Higgs boson marked a triumph of the Standard Model (SM) of particle physics and the Large Hadron Collider (LHC). In the next decades, the LHC and the High-Luminosity LHC (HL-LHC), together with other future colliders in design, will be further exploring the SM and searching for physics beyond that. Among the next critical topics, measuring the vector boson scattering (VBS) processes and especially its longitudinally polarized component (the LL component) is extremely important yet very demanding.

Vector boson scattering is a type of rare SM processes involving pure electroweak interactions. It is sensitive to non-Abelian weak gauge boson interactions and the structure of electroweak symmetry breaking. Typical VBS signatures at hadron colliders include, for example, large di-jet mass and large pseudo-rapidity separation. Many VBS studies have been performed at the LHC, including the discoveries and measurements of $W^{\pm}W^{\pm}$~\cite{Sirunyan:2017ret, Aaboud:2019nmv,Sirunyan:2020gyx}, $W^{\pm}Z$~\cite{Sirunyan:2019ksz, Aaboud:2018ddq,Sirunyan:2020gyx}, $ZZ$~\cite{Aad:2020zbq,Sirunyan:2020alo}, $Z\gamma$~\cite{Khachatryan:2017jub,Aad:2019wpb,CMS:2019iuv,CMS:2021slu}, and $W\gamma$~\cite{Khachatryan:2016vif,Sirunyan:2020azs}. The topic of this paper is the VBS channel of $ZZ\rightarrow 4\ell$ ($\ell=$~electron $e$, or muon $\mu$). While this channel has the advantage of clean final states, it suffers from a low production cross-section, with a small branching-ratio of the $Z$ boson decaying to charged leptons. 

Measuring the longitudinally polarized component of VBS is a critical next step for the field, as it is closely related to the important theoretical property of unitarity restoration through Higgs and possible new physics~\cite{Chang:2013aya,Lee:2018fxj}. 
There have been extensive studies on the LL fraction measurement through exploiting various kinematic observables and techniques like Boosted Decision Tree (BDT) or Deep Learning (DL)~\cite{Searcy:2015apa,Lee:2018xtt,Lee:2019nhm}. However, due to the small yields of this process (typically, the LL fraction is below 10\% of the total VBS), it is quite demanding to distinguish the LL signal from the background. For example, based on the full simulation of samples with the upgraded CMS detector at the 14~TeV HL-LHC~\cite{CMS-PAS-SMP-14-008,CMS-PAS-FTR-18-005,CMS-PAS-FTR-18-014}, the expected significance for an integrated luminosity of 3000\fbinv\ is estimated to be 2.7 and 1.4 standard deviations ($\sigma$), for the LL VBS $W^{\pm}W^{\pm}$ and $ZZ$, respectively. Recently, the first measurement on the LL VBS has been performed for the $W^{\pm}W^{\pm}$ channel by the CMS experiment~\cite{Sirunyan:2020gvn} based on the full Run~2 data of around 137\fbinv. The optimization in the analysis is based on two kinds of BDT, i.e., one to suppress VBS from non-VBS and the other to signify LL from non-LL. The observed significance is only around 1\,$\sigma$. However, it symbols the beginning of the experimental search for the polarized VBS.

\section{Physics processes at the muon collider}
\label{sec:muc}

A muon--muon collider with the center-of-mass energy at the multi-TeV scale has received much-revived interest~\cite{Daniel20} recently, which has several advantages compared with both hadron--hadron and electron--electron colliders~\cite{Mario16,Antonio20,Dario18}. As massive muons emit much less synchrotron radiation than electron beams, muons can be accelerated in a circular collider to higher energies with a much smaller circumference. On the other hand, because the proton is a composite particle, muon--muon collisions are cleaner than proton--proton collisions and thus can lead to higher effective center-of-mass energies. However, due to the short lifetime of the muon, the beam-induced background (BIB) from muon decays needs to be examined and reduced properly. Based on a realistic simulation at $\sqrt{s}=1.5$~TeV with BIB included, Ref.~\cite{Nazar20} found that the coupling between the Higgs boson and the b-quark can be measured at percent level with order\abinv\ of collected data.

At the TeV scale muon collider, VBS can be an interesting topic and ideal motivation. As mentioned in Ref.~\cite{Costantini:2020stv}, a TeV scale muon collider would effectively be a ``high-luminosity weak boson collider'' and offer great opportunities to precisely measure electroweak observables and Higgs coupling.

\begin{figure}[!ht]
    \centering
    \includegraphics[width=0.5\textwidth]{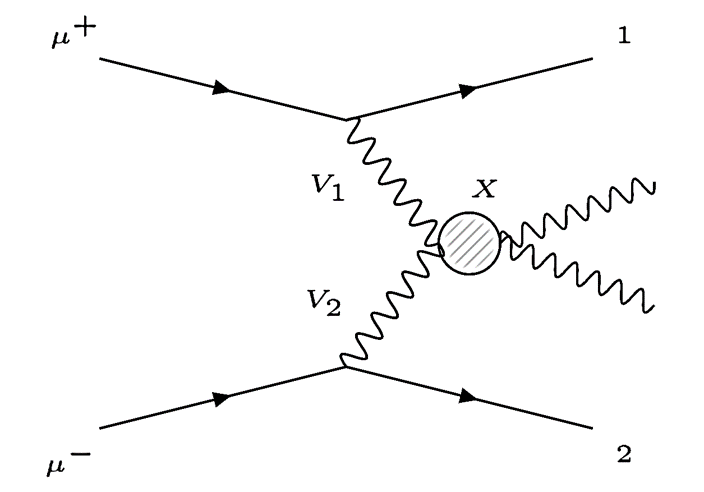}
    \captionsetup{font={small},justification=centering}
    \caption{Example diagram of VBS processes at a muon collider}
    \label{fig:VBS F-diag}
\end{figure}

We target two benchmark scenarios in this study, i.e., a muon--muon collider with a center-of-mass (c.m.) energy of 14~TeV and 6~TeV, each with order ab$^{-1}$ of collected data. The physics processes we consider include (P1) $s$-channel annihilation processes, e.g., $\mu^+\mu^-\rightarrow X= n\, t \bar{t} + m\, V + k\, H $, where $n$, $m$ and $k$ are integers that respectively denote the number of top quark pairs, weak vector bosons $V$, and Higgs bosons $H$, and (P2) VBS processes including the core interactions as $VV\to X$. The example diagram of VBS processes at a muon collider is shown in Fig.~\ref{fig:VBS F-diag}. Following Ref.~\cite{Costantini:2020stv}, VBS processes can be further divided into (P2.1) $W^{+}W^{-}$ fusion with two neutrinos in the final states (symbolized below as $WW$\_VBS), (P2.2) $ZZ/Z\gamma/\gamma\gamma$ fusion with two muons in the final states (symbolized below as $ZZ$\_VBS), and (P2.3) $W^{\pm}Z/W^\pm\gamma$ fusion with one muon and one neutrino in the final states (symbolized below as $WZ$\_VBS). 
\section{Inclusive and polarized $ZZ$ scattering }
\label{sec:vbszz}

In this paper, we focus on the VBS production of $ZZ\rightarrow 4\ell$ at a muon collider, specifically from P2.1 through $W^{+}W^{-}$ fusion. The characteristic signal contains four leptons (electrons or muons) accompanied with the large missing energy $\met$. Accordingly, background processes (from P1, P2.1, P2.2, and P2.3 as listed above) leading to similar final state topology are considered in this study. \par

\begin{figure}[!ht]
    \centering
    \includegraphics[width=0.5\textwidth]{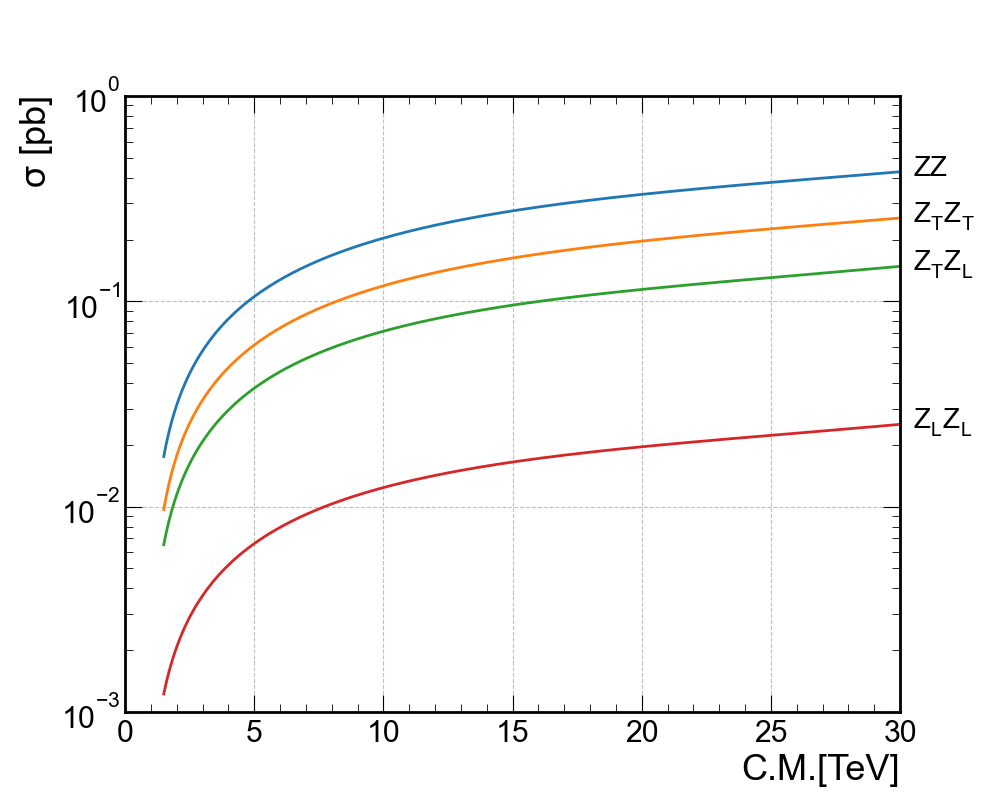}
    \captionsetup{font={small}, justification=raggedright}
    \caption{Cross-section dependence on the center of mass energy of a muon collider, for the LL and other components of VBS $ZZ$ through $W^{+}W^{-}$ fusion.}
    \label{fig:TL_xsec}
\end{figure}

As mentioned before, we are mainly interested in the longitudinally polarized component of the VBS $ZZ$ production, i.e., $Z_LZ_L$ productions through $W^{+}W^{-}$ fusion. A new option to handle the polarization state from initial and final particles has been implemented in \MGMCatNLO~\cite{Alwall:2014hca} from version 2.7.0~\cite{BuarqueFranzosi:2019boy}. This enables the scattering and decay simulations to involve polarized, asymptotic states, and preserves both spin-correlation and off-shell effects, to a good approximation. However, the polarization definition depends on the reference frame. Accordingly, \MGMCatNLO\, provides a setting parameter that allows switching the studied frames. The default is the partonic center-of-mass frame, which is our choice for this work. Notice in the recent CMS analysis~\cite{Sirunyan:2020gvn}, the default reference frame and the VBS frame are both implemented for study, of which the latter has mildly better sensitivity. To further illustrate the difference between the individual polarization states, we plot in Fig.~\ref{fig:TL_xsec} the cross-section dependence on the collision energy, for VBS productions of $ZZ$, $Z_LZ_L$, the transversely polarized component $Z_TZ_T$, as well as $Z_TZ_L$ through $W^{+}W^{-}$ fusion at a muon collider.

\section{Simulation and analysis framework}
\label{sec:simulation}

Both signal and background events are simulated with \MGMCatNLO, then showered and hadronized by \pythia8~\cite{Sjostrand:2014zea}. The final state jets are clustered using \fastjet~\cite{Cacciari:2011ma} with the \kt~\cite{Cacciari:2008gp} algorithm at a fixed cone size of $R_{\rm jet}=0.5$. We used \delphes~\cite{deFavereau:2013fsa} version 3.0 to simulate detector effects with the default card for the muon collider detector~\cite{mucard}. The simulated backgrounds are summarized in Table~\ref{tab:table1}. %
\begin{table}[htbp]
    \centering
    \caption[]{Summary of the background processes at a muon collider considered in this study.}
    \label{tab:table1}
    \begin{minipage}{0.8\textwidth}
    \begin{tabularx}{\textwidth}{p{2.65cm}<{\centering}X<{\centering}}
     \hline\hline
        SM process type&Selected background\\
    \hline
        P1: $s$-channel &$ZZ$, $WWZ$\\
        P2.1: $WW$\_VBS &$H$, $HZ(Z)$, $HWW$, $HH$, $WWZ$, $ZZZ$, $Z_TZ_{T,L}$, $t\bar{t}Z$\\
        P2.2: $ZZ$\_VBS &$H$, $WW$, $t\bar{t}$, $4e$, $2e2\mu$, $4\mu$\\
        P2.3: $WZ$\_VBS &$WZ$, $WZH$, $WH$, $WWW$, $WZZ$\\
     \hline\hline
    \end{tabularx}
    \end{minipage}
\end{table}
Events are generated corresponding to a muon collider with collision energies of 14 (6)~TeV and integrated luminosity of 20 (4)\abinv. 

We describe the selection criteria as below. First, events must include exactly four leptons with transverse momentum $\pt>20$~GeV and absolute pseudo-rapidity $|\eta|<2.5$, and satisfy the lepton flavor and charge requirements from the $Z$ boson decay, i.e., the leptons should fall into one of the three flavor types ($4e$, $4\mu$, $2e2\mu$) with exactly two positive and two negative charges. For the type $2e2\mu$, the di-electron and di-muon pair must have opposite charges separately. Furthermore, we veto events with jets cleaned from the four leptons with a distance of $\Delta R=\sqrt{\Delta^2{\eta}+\Delta^2{\phi}}>0.5$. We then cluster the selected four leptons into two reconstructed ``$Z$ bosons'' (labeled $Z_1$ and $Z_2$ in the descending order of transverse momenta) using the following algorithm:
\begin{itemize}
    \vspace{0.15cm}
    \item Construct the combinatorics of the leptons $l_1^+l_2^+l_3^-l_4^-$ into lepton pair candidates ($l_1l_3,\,l_2l_4$) and ($l_1l_4,\,l_2l_3$),
    \vspace{0.15cm}
    \item Calculate $\Delta M^2=(M_{Z_1'}-M_Z)^2+(M_{Z_2'}-M_Z )^2$,
    \vspace{0.15cm}
    \item If $\Delta M_{13,24}^2>\Delta M_{14,23}^2$, choose the lepton pair ($l_1l_4,\,l_2l_3$), and vice versa,
    \vspace{0.15cm}
    \item For the flavor type $2e2\mu$, adopt the lepton paring $Z_1 \rightarrow e^+e^-$, $Z_2\rightarrow \mu^+\mu^-$.
    \vspace{0.15cm}
\end{itemize}

\begin{figure}[!ht]
    \centering
    \includegraphics[width=0.85\textwidth]{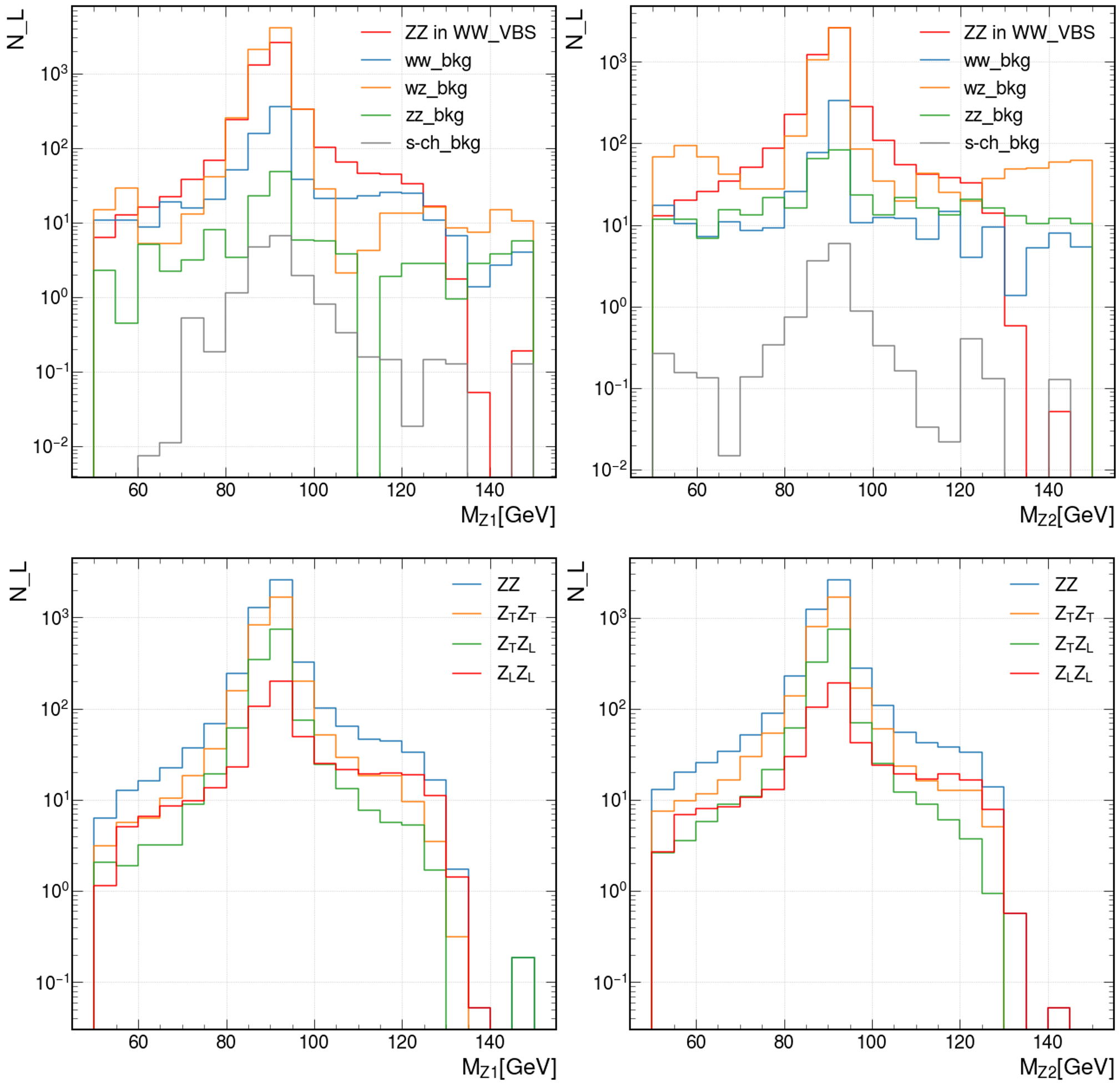}
    \captionsetup{font={small}, justification=raggedright}
    \caption{The invariant mass distributions of $Z_1$ (left) and $Z_2$ (right) for VBS $ZZ$ through WW fusion and its various polarization fractions, compared with all the backgrounds, corresponding to a 14~TeV muon collider with 20\abinv of data collected. The upper panel compares the VBS $ZZ$ signal with four background components. The lower panel compares the kinematics for different polarization fractions of the VBS $ZZ$ process.}
    \label{fig:MZ_compare}
\end{figure}

It is worth mentioning that the four leptons in some background events are not originated from the two $Z$ bosons. As a result, the reconstructed ``$Z$ boson'' in those events may behave differently from the real $Z$ boson, thus can be exploited to distinguish our signal from the background. Moreover, even for leptons produced from the decay of a real $Z$ boson, we observe discrepancies in their distributions among different polarization states. Fig.~\ref{fig:MZ_compare} shows the invariant mass distributions of $Z_1$ and $Z_2$ for VBS $ZZ$ through $WW$ fusion and its various polarization fractions, compared with all the backgrounds. Given the intrinsic shape discrepancies, we train a BDT model for the better discrimination of the signal and all background components, using the kinematic observables of the four leptons and two ``$Z$ bosons'' as input features. We elaborate on the details in the next section.


\section{Analysis results}
\label{sec:results}
\par For the implementation of the BDT, we shuffle the signal and background events and define the training and test sets with the event ratio of $2:1$. The BDT with 200 trees and a maximum depth of 5 is trained. We apply the per-event weight during the training to account for the cross-section difference among the processes. The weight is defined by
\begin{equation}
    n_{L_X}=\sigma_XL/N_{G_X},
\end{equation}
where $\sigma_X$ denotes the cross-section of a process, $L$ denotes the default target luminosity in this study as 20 (4)\abinv\ for a 14 (6) muon collider, and $N_{G_X}$ denotes the generated number of events. The total signal yields are reweighted to match that of the total background during training for the robustness of the trained model. The input features, namely reconstructed kinematics of each event used for training, are listed as follows and summarized in Table~\ref{tab:table2}. In this analysis, up to 37 kinds of event kinematics are used for the BDT model. 

\begin{itemize}
\vspace{0.15cm}
\item $(\pt,\,\eta,\,\phi)$ of the four leptons (note that due to the small mass of the leptons, we set the masses to zero in the calculations),
\vspace{0.15cm}
\item $(\pt,\,\eta,\,\phi,\,m_{\mathrm{inv}})$ of the 2 clustered ``$Z$ bosons'',
\vspace{0.15cm}
\item $(\pt,\,\eta,\,\phi,\,m_{\mathrm{inv}})$ of the summed 4-momenta of the four leptons,
\vspace{0.15cm}
\item $(\pt,\,\eta,\,\phi)$ of the missing energy \met,
\vspace{0.15cm}
\item $(\Delta \eta,\,\Delta \phi,\,\Delta R)$ between the 2 ``$Z$ bosons'' and the 2 lepton pairs from $Z_1$ and $Z_2$,
\vspace{0.15cm}
\item The flavor types of the 4 leptons $(4e,\,4\mu,\,2e2\mu)$, encoded as $(1,\,-1,\,0)$ respectively.
\end{itemize}

\begin{table}[htbp]
    \centering
    \caption[]{Summary of features used for the BDT training.}
    \label{tab:table2}
    \begin{minipage}{0.8\textwidth}
    \begin{tabularx}{\textwidth}{X<{\centering}X<{\centering}p{2cm}<{\centering}}
     \hline\hline
     Objective&Features&Number of features\\
    \hline
        Each lepton&$(\pt,\,\eta,\,\phi)$&12\\
        Each ``$Z$ boson"&$(\pt,\,\eta,\,\phi,\,m_{\mathrm{inv}})$&8\\
        Four leptons combined&$(\ptfl,\,\eta_{4l},\,\phi_{4l},\,m_{4l})$&4\\
        \met&$(\pt,\,\eta,\,\phi)$&3\\
        Between two Z bosons&$(\Delta \eta,\,\Delta \phi,\,\Delta R)$&3\\
        Between $2\ell$ of $Z_1$&$(\Delta \eta,\,\Delta \phi,\,\Delta R)$&3\\
        Between $2\ell$ of $Z_2$&$(\Delta \eta,\,\Delta \phi,\,\Delta R)$&3\\
        Lepton flavor type&$(1,\,-1,\,0)$ for $(4e,\,4\mu,\,2e2\mu)$&1\\
        Total:&&37\\
     \hline\hline
    \end{tabularx}
    \end{minipage}
\end{table}

\begin{figure}[!ht]
    \centering
    \subfloat[Top 10 Kin. Importance]{\label{sfig:FeatureImportance}\includegraphics[height=0.3\textwidth]{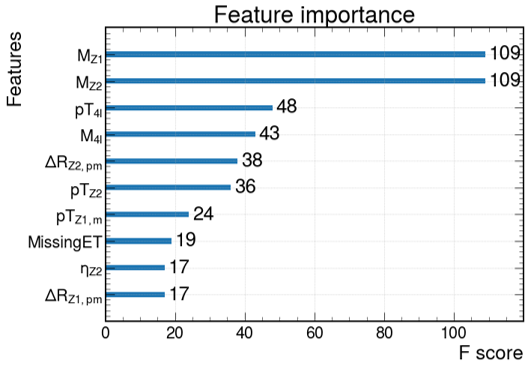}}\qquad 
    \subfloat[Over Training Test]{\label{sfig:OTTest}\includegraphics[height=0.3\textwidth]{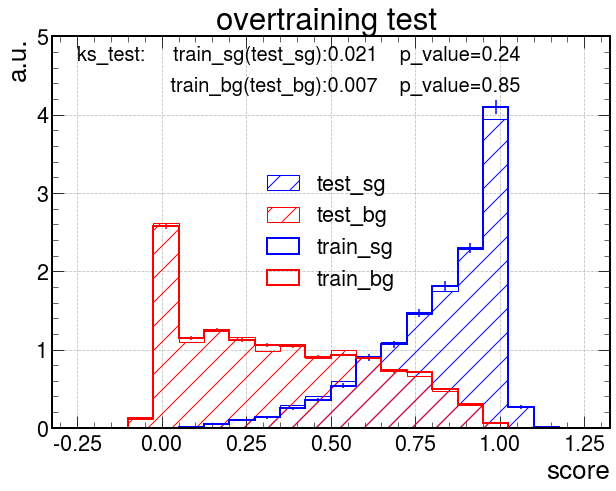}}\qquad 
    \caption{BDT training results for the LL VBS $ZZ$ analysis at a 14~TeV muon collider. The left panel \ref{sfig:FeatureImportance} shows the importance ranking for the top 10 features; the right panel \ref{sfig:OTTest} shows the Kolmogorov-Smirnov test results and score distributions for signal and background in the training and test sets, respectively. }
    \label{fig:features}
\end{figure}

The results of the BDT is shown in Fig.~\ref{fig:features}. In Fig.~\ref{sfig:FeatureImportance}, we show the feature importance ranking of the top 10 variables in the BDT model, from which we can easily conclude that the invariant masses of the two reconstructed ``$Z$ bosons'' are of vital importance in discriminating the signal and backgrounds. Fig~\ref{sfig:OTTest} provides p-values from Kolmogorov-Smirnov test and the BDT score distributions for the signal and background in the training and test sets, as proof of no over-training in the BDT model.  

\begin{figure}[!ht]
    \centering
    \subfloat[ROC Curve of Trained BDT Model]{\label{sfig:ROC}\includegraphics[height=0.4\textwidth]{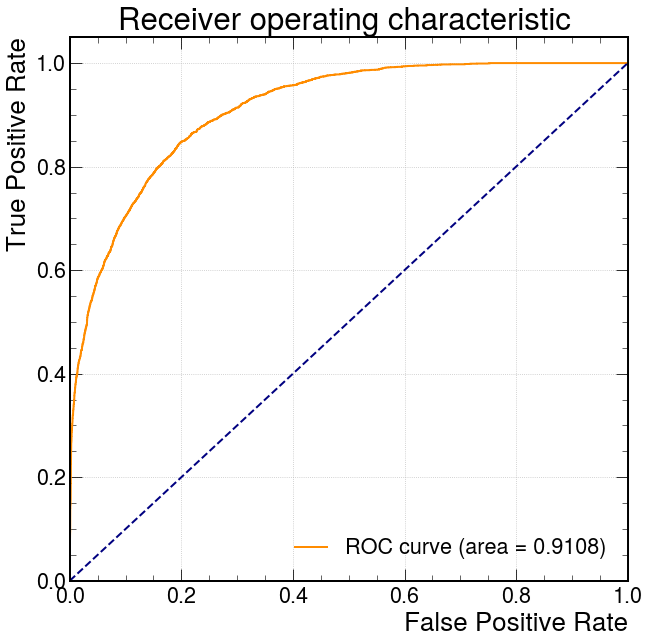}}\qquad 
    \subfloat[Thresholds on BDT Score vs. Corresponding Significance]{\label{sfig:Envelope}\includegraphics[height=0.4\textwidth]{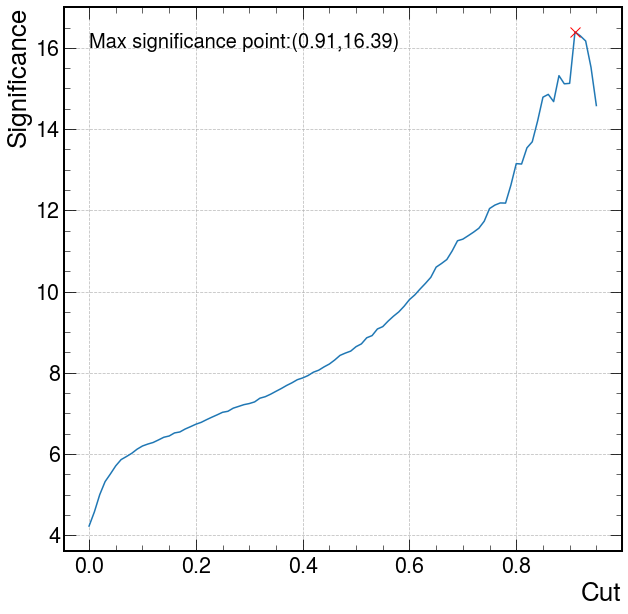}}
    \captionsetup{font={small},justification=centering}
    \caption{ROC curve and significance dependence on the BDT cut, corresponding to a 14~TeV muon collider with 20\abinv\ of data collected. }
    \label{fig:ROC&Sig-Cut}
\end{figure}

The receiver operating characteristic (ROC) curve of the trained model is then studied from the test set. The significance is also calculated at different BDT selection thresholds, using the formula
\begin{equation}
    S=\sqrt{2(s+b)\mathrm{ln}(1+\frac{s}{b})-2s},
\end{equation}
where $s$ ($b$) represents the weighted signal (background) yields after requiring the BDT score to be greater than a certain value. The ROC curve and the significance dependence on the BDT score cut are shown in Fig.~\ref{fig:ROC&Sig-Cut}.  

\begin{figure}[!ht]
    \centering
    \subfloat[Results corresponding to different random datasets]{\label{sfig:Rdm}\includegraphics[height=0.4\textwidth]{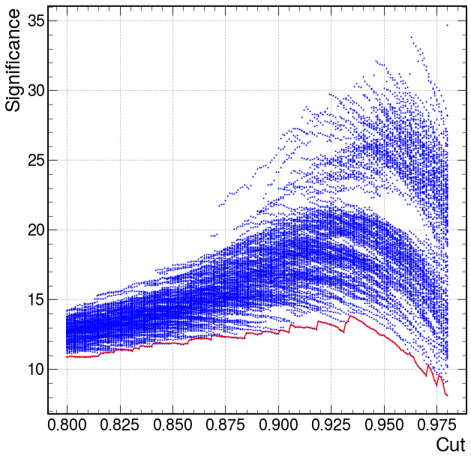}}\qquad 
    \subfloat[Lower envelope of~Fig.\ref{sfig:Rdm}]{\label{sfig:Envelope}\includegraphics[height=0.4\textwidth]{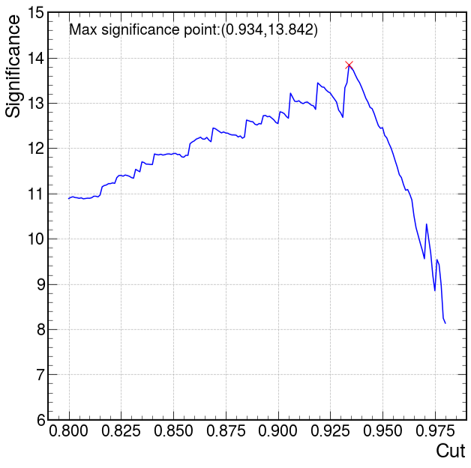}}
    \captionsetup{font={small},justification=centering}
    \caption{Training results with the use of 150 different random seeds. The left plot shows the significance dependence on the BDT cut for all 150 scenarios. It should be noted that we pause the scanning when the backgrounds yield zero, which results in the straight horizontal line near the end of scanning. The right plot shows the lower envelope of the significance shape.}
    \label{fig:150seeds}
\end{figure}

It is worth pointing out that, events from some processes with large cross-sections, such as the $WZH,WH$ and $WZZ$ from $WZ\_VBS$ (conventions see Table\ref{tab:table1}), have relatively large weight ($n_{L_X}$) after scaling to target luminosity as the consequence of the intractable complexity in event generation and the limited CPU resources. Events with big $n_{L_X}$ play an important role when we randomly assign MC events into separate datasets for training and testing respectively and lead to a large fluctuation on BDT scores as shown in Fig.~\ref{fig:ROC&Sig-Cut}. Although in most situation tighter cut on BDT score leads to better significance, fluctuation of achieved significance corresponding to BDT cut approaching 1 confirms us to adopt a conservative strategy and pause the scanning of threshold at 0.95. To further validate the robustness of our results, we split our simulated events to training and testing sets with 150 random configuration and train 150 same BDT models respectively. Our target is to extract a reasonable significance with the collection of all 150 trained models. Results for this validation are illustrated in Fig.~\ref{fig:150seeds}. 

In the left panel~Fig.\ref{sfig:Rdm}, we show the training results with the  150 random dataset configuration. The right plot~Fig.\ref{sfig:Envelope} shows the lower envelope of the significance as a function of the BDT cut, in these 150 training sessions. We observe that the optimal cut value is around 0.93, with the corresponding significance of around 14\,$\sigma$. Such a result is the most conservative estimation of the LL VBS $ZZ$ signal significance for a 14~TeV muon collider.
 
We also make a comparison between the BDT method and the traditional cut-base method. The cut-flow results are listed in Table~\ref{tab:table3}, where $\Delta R_{Z2,pm}$ is defined as the $\Delta R$ between the two leptons forming $Z_2$. We use the selections listed in Table~\ref{tab:table3} to filter all four-lepton events. The selections are optimized to reach a higher significance. Comparing the maximum significance reachable by the BDT and the cut-base selection method, we see a substantial gain from the BDT method in the distinction between signal and background. Besides, we derive the integrated luminosity where a maximum significance of 5\,$\sigma$ is reached by $L'=(5^2\big/14^2)\,L\approx 3000\fbinv$. We conclude that the LL VBS $ZZ$ process can be eventually discovered on a 14~TeV muon collider with 3000\fbinv\ of collected data, resulting from a conservative estimation by the BDT method.

\begin{table}[htbp]
    \centering
    \caption[]{The cut-flow table and the corresponding significance based on the cut-based method. The selections are optimized to achieve higher significance.}
    \label{tab:table3}
    \begin{minipage}{0.85\textwidth}
    \begin{tabularx}{\textwidth}{p{8cm}<{\centering}X<{\centering}X<{\centering}X<{\centering}}
     \hline\hline
        cuts&$s$&$b$&$S\,[\sigma]$\\
    \hline
        $70\mathrm{GeV}<M_{Z1}$, $M_{Z2}<140\mathrm{GeV}$&476.5&6592.1&5.8\\
        $70\mathrm{GeV}<M_{Z1}$, $M_{Z2}<140\mathrm{GeV}$, $\Delta R_{Z2,pm}<0.4$&238.1&1165.9&6.8\\
        $70\mathrm{GeV}<M_{Z1}$, $M_{Z2}<140\mathrm{GeV}$, $\Delta R_{Z2,pm}<0.4$,&213.5&424.9&9.6\\
        $\ptfl<300\mathrm{GeV}$&&&\\
        $70\mathrm{GeV}<M_{Z1},M_{Z2}<140\mathrm{GeV}$, $\Delta R_{Z2,pm}<0.4$,&147.8&158.1&10.4\\
        $\ptfl<300\mathrm{GeV}$, $\met<140\mathrm{GeV}$&&&\\
     \hline\hline
    \end{tabularx}
    \end{minipage}
\end{table}

\begin{figure}[!ht]
  \centering
  \subfloat[$\Delta R_{{Z_2},pm}$, $\sqrt{s}=14~\text{TeV}$]{\label{sfig:dr14TeV}\includegraphics[height=0.4\textwidth]{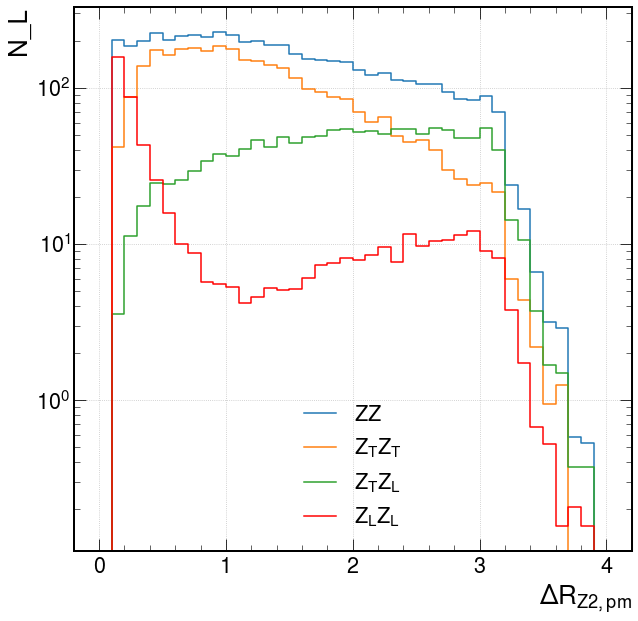}}\qquad
  \subfloat[$\Delta R_{{Z_2},pm}$, $\sqrt{s}=6~\text{TeV}$]{\label{sfig:dr6TeV}\includegraphics[height=0.4\textwidth]{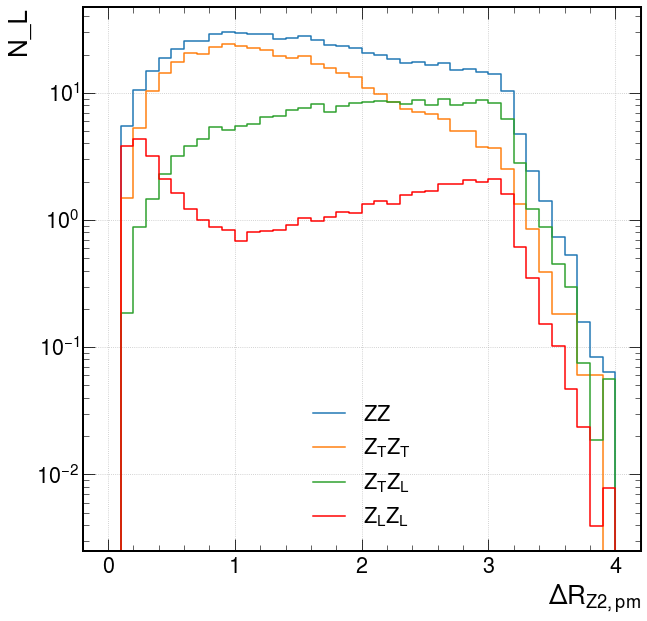}}\\
  \subfloat[$\ptfl$, $\sqrt{s}=14~\text{TeV}$]{\label{sfig:pt4l14TeV}\includegraphics[height=0.4\textwidth]{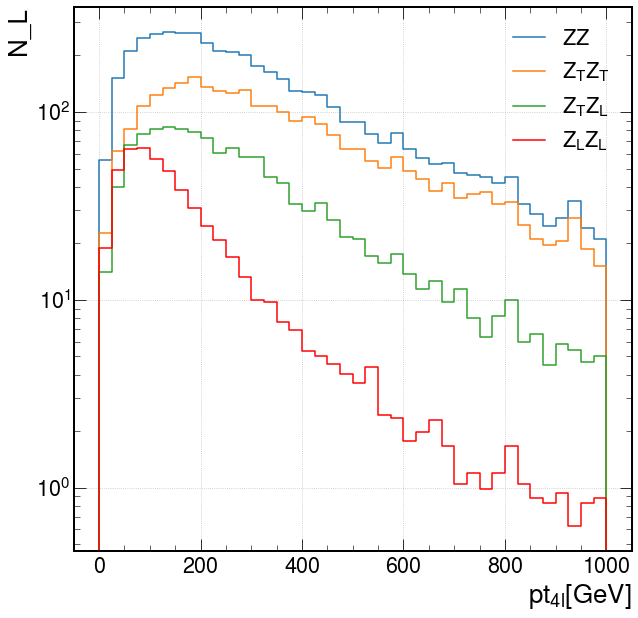}}\qquad
  \subfloat[$\ptfl$, $\sqrt{s}=6~\text{TeV}$]{\label{sfig:pt4l6TeV}\includegraphics[height=0.4\textwidth]{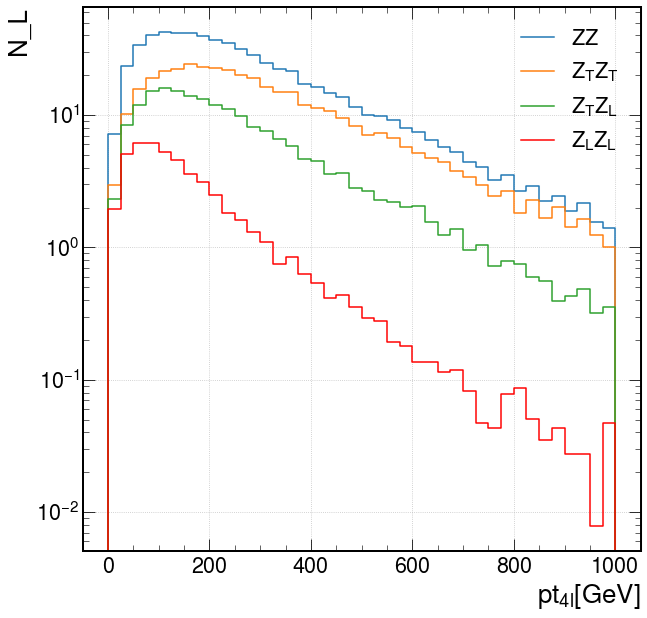}}
  \caption{Distributions of $\Delta R_{{Z_2},pm}$ (\ref{sfig:dr14TeV}~\&~\ref{sfig:dr6TeV}), defined as the $\Delta R$ between the two leptons forming $Z_2$, and $\ptfl$ (\ref{sfig:pt4l14TeV}~\&~\ref{sfig:pt4l6TeV}) from the VBS $ZZ$ process simulated at a 14~TeV (\ref{sfig:dr14TeV}~\&~\ref{sfig:pt4l14TeV}) and a 6~TeV (\ref{sfig:dr6TeV}~\&~\ref{sfig:pt4l6TeV}) muon collider. }
  \label{fig:14TeV_vs_6TeV}
\end{figure}

We also show the results of our study at a 6~TeV muon collider. With the same BDT study performed as above, the expected significance for LL VBS $ZZ$ is found to be around 2\,$\sigma$ with 4\abinv, which is much weaker than the 14~TeV case. However, it is already comparable with the HL-LHC estimation~\cite{CMS-PAS-FTR-18-014}. We explain the details of the differences between a 6 and 14~TeV muon collider as follows. (1) The $N_L$ of our signal after the pre-selection at $\sqrt{s}=6$~TeV is only one-tenth of that at $\sqrt{s}=14$~TeV. (2) Compared to $\sqrt{s}=14$~TeV, the different polarization states of the $ZZ$ process in $WW$\_VBS at a 6~TeV muon collider are more similar in their kinematic distributions and thus difficult to be distinguished. For a better illustration, we plot the distributions of $\Delta R_{pm,Z2}$ and $\ptfl$ at two different collision energies in Fig.~\ref{fig:14TeV_vs_6TeV}. Comparing the left two sub-figures with the right ones, we can see that in the region where the values of two selected observables approach 0, the signal process at a 14~TeV muon collider has a more prominent distribution than at a 6~TeV muon collider. As a result, it becomes more difficult to ensure the number of the signal remains a considerable size when reducing the background yields.


Finally, we want to comment on the effects of lepton isolation, which represents the requirement on the detector granularity of a muon collider. In above analysis, we follow the \delphes~card for muon colliders~\cite{mucard}, where the lepton isolation cone is set to 0.1 by default. However, from Fig.~\ref{fig:14TeV_vs_6TeV}, one can find LL VBS $ZZ$ peak sharply at low $\Delta R_{{Z_2},pm}$ region~\footnote{We have checked that at the Z boson's rest frame, leptons from Z decay tend to be perpendicular to the Z boson's fly direction in the lab frame. However, at the 14 TeV muon collider, Z bosons are quite energetic and the decay products are boosted to be closer, especially for the longitudinal mode.}, and if one cuts on $\Delta R_{{Z_{1,2}},pm}$ to be larger than e.g. 0.2~\cite{LClep,Yonamine:2011jg}, the significance can drop to 6 $\sigma$ as shown in Table~\ref{tab:table4}. This clearly points out to the needs of detector optimization, in order to enlarge the gain of measuring LL VBS at a muon collider.

\begin{table}[htbp]
    \centering
    \caption[]{The cut-flow table and the corresponding significance when $\Delta R_{{Z_{1,2}},pm}>0.2$.}
    \label{tab:table4}
    \begin{minipage}{0.85\textwidth}
    \begin{tabularx}{\textwidth}{p{8cm}<{\centering}X<{\centering}X<{\centering}X<{\centering}}
     \hline\hline
        cuts&$s$&$b$&$S\,[\sigma]$\\
    \hline
        $\Delta R_{{Z_{1,2}},pm}>0.2$&334.3&14331.2&2.8\\
        $0.2<\Delta R_{{Z_{1}},pm}<0.8$, $0.2<\Delta R_{{Z_{2}},pm}<0.5$&108.7&1007.6&3.4\\
        $0.2<\Delta R_{{Z_{1}},pm}<0.8$, $0.2<\Delta R_{{Z_{2}},pm}<0.5$,&100.0&695.4&3.7\\
        $60\mathrm{GeV}<M_{Z1},M_{Z2}<130\mathrm{GeV}$&&&\\
        $0.2<\Delta R_{{Z_{1}},pm}<0.8$, $0.2<\Delta R_{{Z_{2}},pm}<0.5$,&97.0&400.7&4.7\\
        $60\mathrm{GeV}<M_{Z1},M_{Z2}<130\mathrm{GeV}$, $\ptfl<500\mathrm{GeV}$&&&\\
        $0.2<\Delta R_{{Z_{1}},pm}<0.8$, $0.2<\Delta R_{{Z_{2}},pm}<0.5$,&61.7&90.2&5.9\\
        $60\mathrm{GeV}<M_{Z1},M_{Z2}<130\mathrm{GeV}$, $\ptfl<500\mathrm{GeV}$,&&&\\
        $M_{4l}<3000\mathrm{GeV}$, $\met<180\mathrm{GeV}$&&&\\
     \hline\hline
    \end{tabularx}
    \end{minipage}
\end{table}
\section{Outlook and conclusions}
\label{sec:conclusions}

Measuring longitudinally polarized vector boson scattering in the $ZZ$ channel is a promising way to investigate unitarity restoration with the Higgs mechanism and to search for possible new physics. However, at the LHC, it demands the end of the HL-LHC lifetime luminosity (i.e., 3000\fbinv) and advanced data analysis technique in order to reach the discovery threshold, due to the small production rates. We show that there could be great potential for the discovery of this signal process at future linear colliders. In this paper, we perform a Monte Carlo study and examine the projected sensitivity of longitudinally polarized $ZZ$ scattering at TeV scale muon colliders. With a conservative estimation using the BDT technique, we find that a 5 standard deviation discovery can be achieved at a 14~TeV muon collider, with 3\abinv\ of data collected. The result outperforms the traditional cut-based analysis strategy. The study is also performed on a 6~TeV muon collider and shows that 2 standard deviations can be reached with around 4\abinv\ of data. Although less discriminating power, w.r.t. the 14~TeV muon collider, is observed to distinguish the longitudinal component of VBS $ZZ$ from the backgrounds, the 2\,$\sigma$ significance already surpasses the estimated sensitivity at the HL-LHC, showing good potentials to reach the first VBS $Z_LZ_L$ discovery on a muon collider.
The effect from lepton isolation and detector granularity is also discussed, which shows to be larger at higher energy muon colliders, as the leptons from longitudinally polarized Z decays tend to be closer.


\begin{acknowledgments}
This work is supported in part by the National Natural Science Foundation of China under Grants No. 12075004 and No. 12061141002, by MOST under grant No. 2018YFA0403900.
\end{acknowledgments}

\appendix
\label{sec:appendix}


\end{document}